\documentclass[12pt]{article}
\usepackage{setspace}
\usepackage{amsmath,amsfonts}
\usepackage{graphicx}
\usepackage{color}
\usepackage{times}
\usepackage{bm}
\usepackage{natbib}
\usepackage[plain,noend]{algorithm2e}
\usepackage{url}

\usepackage{tikz}

\newtheorem{theorem}{Theorem}[section]

\usetikzlibrary{backgrounds}
\usetikzlibrary{decorations}
\usetikzlibrary{decorations.pathmorphing}
\usetikzlibrary{decorations.pathreplacing}
\usetikzlibrary{decorations.shapes}
\usetikzlibrary{decorations.text}
\usetikzlibrary{decorations.markings}
\usetikzlibrary{decorations.fractals}
\usetikzlibrary{decorations.footprints}

\newenvironment{comment}%
 {\begin{quote}\color{red}\begin{sffamily}}%
 {\end{sffamily}\color{black}\end{quote}}

\newcommand{\argmin}{\text{arg min}} 
\newcommand{\E}{\ensuremath{\mathbb{E}}} 
\newcommand{\sign}{\textrm{sign}}			
\newcommand{\rset}{\ensuremath{\mathbb{R}}}
\newcommand{\nset}{\ensuremath{\mathbb{N}}}

\def\1{{\mbox{\rm 1\hspace{-0.25em}I}}}  
\newcommand{\N}{\mathcal{N}} 
\newcommand{\gdO}{\mathcal{O}} 			

\newcommand{\betabf}{\text{\mathversion{bold}{$\beta$}}}

\newcommand{\mubf}{\text{\mathversion{bold}{$\mu$}}}
\newcommand{\varepsilonbf}{\text{\mathversion{bold}{$\varepsilon$}}}
\newcommand{\pibf}{\text{\mathversion{bold}{$\pi$}}}

\newcommand{\tbf}{\mathbf{t}}

\newcommand{\Ybf}{\mathbf{Y}}

\newcommand{\Wbf}{\mathbf{W}}

\newcommand{\dbf}{\mathbf{d}}

\newcommand{\gbf}{\mathbf{g}}

\newcommand{\zero}{\text{\mathversion{bold}{$0$}}}

\newcommand{\Ebf}{\mathbf{E}}
\newcommand{\Gbf}{\mathbf{G}}

\def\1{1\!{\rm l}}

\title{Minimax wavelet estimation for multisample heteroscedastic non-parametric regression}

\author{Madison Giacofci$^{\dag}$, Sophie Lambert-Lacroix$^{\circ}\footnote{Corresponding author}$, Franck Picard$^{\star}$ \\ \\ 
  $^\dag$ Laboratoire LJK, Universit\'e de Grenoble et CNRS \\
  $^\circ$ UJF-Grenoble 1/CNRS/UPMF/TIMC-IMAG UMR 5525 \\
  $^\star$ LBBE, UMR CNRS 5558 Universit\'{e} Lyon 1
}

\begin{document}

\maketitle


\begin{abstract}
The problem of estimating the baseline signal from multisample noisy curves is investigated. We consider the functional mixed effects model, and we suppose that the functional fixed effect belongs to the Besov class. This framework allows us to model curves that can exhibit strong irregularities, such as peaks or jumps for instance. The lower bound for the $L_2$ minimax risk is provided, as well as the upper bound of the minimax rate, that is derived by constructing a wavelet estimator for the functional fixed effect. Our work constitutes the first theoretical functional results in multisample non parametric regression. Our approach is illustrated on realistic simulated datasets as well as on experimental data.
\end{abstract}

{\bf keywords.-}functional mixed effects models; minimax risk; Besov class; wavelet estimator;

\newpage

\section{Introduction}

Functional data analysis has gained increased attention in the past years, in particular in high-throughput biology with the use of mass spectrometry. In this field, the signal is a spectrum whose peaks provide information regarding the protein content of biological samples. A new challenge in functional data analysis is the availability of multisample data for which functional ANOVA has become the appropriate framework. More specifically for spectrometry data, it is now well accepted that the noise corrupting the signal can be divided into a technical white noise added to an important inter-individual variability \citep{EOT09}. In this case, the usual non-parametric regression framework (a deterministic trend corrupted by a random noise) is no longer appropriate since it does not account for heteroscedastic noise structure. Functional mixed effects models \citep{AS07} appear to be a powerful framework to handle these data, as others, and we focus here on the estimation of the baseline signal.

In practice, a trivial averaging procedure is often used to get an estimate of the baseline signal, but it has both a poor convergence rate and a finite sample performance. \cite{AS05} proposed an approach for baseline estimation based on empirical wavelet coefficients of the observed data. Unfortunately the convergence of their estimator is not theoretically assessed, and more broadly, there is a general lack of theoretical results on functional estimators in functional mixed models, despite their increasing importance in practice \citep{MC06,MBH08}.

In this work we propose a minimax estimator of the baseline signal, based on the empirical wavelet coefficients of the observed data. The functional fixed effect is assumed to belong to the Besov class, which allows us to model curves that can exhibit strong irregularities, such as peaks in mass spectrometry data. We construct the lower bound for the $L_2$ minimax risk. This convergence rate is the same as in the classical non parametric setting but with an additional approximation error term. Then, we propose a wavelet estimator that achieves near optimal rate of convergence (within a logarithmic factor in sample size). Through simulation studies, we show that our approach outperforms the  approach proposed by \cite{AS05}. We also propose a new thresholding procedure based on the Stein Unbiased Risk Estimate (SURE) \citep{St81}, combined with the SCAD thresholding \citep{AF01}. This leads to improved performance for the baseline signal estimation. 

This article is organized as follows. Section 2 presents the heteroscedastic model and the theoretical properties of our minimax estimator (lower and upper bounds). In particular we show how classical rates are modified in the presence of replicates along with inter-individual variability. Most of all, our work constitutes the first theoretical functional results in heteroscedastic multisample non-parametric regression. Several thresholding strategies are considered in Section 3, where we provide a new SURE-based procedure. Section 4 is devoted to the numerical experiments, and the procedure is illustrated on an experimental dataset. Technical proofs are provided in the Appendix.

\section{Heteroscedastic nonparametric regression model and theoretical properties \label{model description}}

\subsection{Functional model}

We observe $N$ curves $Y_i(\cdot)$, for $i =1,\ldots, N,$ over $M$ equally spaced time points $\tbf = (t_1,\hdots,t_M)$ in $[0,1]^M$, with $M=2^J$ for some integer $J$. In the general functional setting we consider a functional modeling (as in \cite{AS07}) for the observed signal of the $i$th individual:
\begin{equation}
\label{FunctionalModel}
  Y_i(t_j) = \mu(t_j) + E_i(t_j), \quad \forall i = 1,\ldots,N,\quad \forall j = 1,\ldots,M,
\end{equation}
where $E_i(\cdot)$, for $i=1,\ldots,N,$ are stochastically independent random functions that are modeled as realizations of zero-mean Gaussian processes
with parametrically structured covariances modeled in the wavelet domain (see Section \ref{ssWE}). We define $\mu$ to be the main functional fixed effect characterizing a population average profile. In the following, we will denote by $\mathbf{Y}_i = (Y_i(t_1),\hdots,Y_i(t_M))$, $i = 1,\ldots,N$, the vector of observations on the time grid, and similarly by $\mubf$ and $\Ebf_i$, $i=1,\ldots,N$, respectively the vector of the fixed effect and the noise terms, observed on the discrete time grid. 

This modeling allows us to account for functional mixed effects models by decomposing $E_i(t)$ in a sum of two independent processes $E_i(t_j) = U_i(t_j)+\epsilon_{ij}$, where $\epsilon_{ij}$ are independent and identically distributed Gaussian random variables  with zero-mean and constant variance; $U_i(t)$ is a centered Gaussian process standing for subject-specific functional deviations. In \cite{AS05}, the authors introduce similar model although the variance of the process  $U_i(t)$ is constant with respect to positions $t_j$.

\subsection{Minimax approach}
In what follows we suppose that $\mu$ belongs to the Besov class ${\cal F}={\cal F}(s,p,q,L)$ (see Section \ref{ssWE} for a proper definition), a set of compactly supported functions (on $[0,1]$) with a bounded Besov space norm (by $L$). Such a set allows to model curves that can exhibit strong irregularities, such as peaks or jumps for instance. The notion of regularity is at the core of the functional setting which makes inhomogeneous Besov spaces a privileged tool for irregular function analysis. These spaces allow the fine definition of the regularity $s$ of a function along with its derivatives lying in $L^p([0,1])$ while bringing a correction $q$ to this regularity. For a detailed review of Besov spaces and their properties, we refer the reader to the books of \cite{HKPT98} or \cite{DL93}. 

Our goal is to recover the main functional effect $\mu$ from noisy observations. An originality of our approach is to consider multiple, say $N$, individuals, which constitute available replicates to estimate the main fixed effect. To derive our estimator, we propose to use the so-called minimax approach. In this setup the risk of an estimator $\widehat{\mu}_{N,M}$ is defined by $\mathbb{E} \big( \| \widehat{\mu}_{N,M}-\mu \| \big)$, with $\|\cdot\|$ being a functional norm or a semi-norm. Then the so-called \textit{minimax} estimator, denoted by $\widehat{\mu}^*_{N,M}$, is the minimizer of the maximal risk on class ${\cal F}$ over the set of all estimators: $$ \mathcal{R} (\widehat{\mu}_{N,M},{\cal F})=\sup_{\mu \in {\cal F}}\mathbb{E} \big( \| \widehat{\mu}_{N,M}-\mu \| \big).$$ Thus the challenge is to propose an optimal minimax estimator $\widehat{\mu}^*_{N,M}$, and to derive its associated risk $\mathcal{R}^*_{N,M} ({\cal F}) = \mathcal{R} (\widehat{\mu}^*_{N,M},{\cal F})$, also referred to as the minimax risk.

The construction of minimax estimators on the Besov classes is well known when only one replicate is available (see \cite{HKPT98}). When errors are measured with a $L_r$-norm ``sharper'' than the norm of the functional class $p$, wavelet-based thresholding estimators can significantly outperform linear projection estimates. The rate of convergence depends on $r$, $p$ and $s$ with two zones: the regular zone with usual rate $M^{-s/(2s+1)}$ and the sparse zone with a slower rate of convergence. However, this rate is not known when replicates are available ($N>1$). In this work we establish this risk for $r=2$ (we will denote this norm by $\|\cdot\|_2$) and for the Besov class ${\cal F}$ with usual constraints  $p \geq 1$, $q \geq 1$ and $s \geq 1/p$. That leads to consider the regular zone since, in this case, we have $s' = s-1/p+1/2 >0$ (see \cite{HKPT98}). In order to establish the minimax risk, we first give its lower bound and secondly we propose an estimator that achieves a near optimal rate of convergence. In this context, the near-optimality means that the minimax rate is attained within a logarithmic factor in sample size $M$.

\subsection{Lower bound for the minimax risk}

One of the main contributions of this paper is to derive the asymptotic lower and upper bounds for $\mathcal{R}_{N,M} ({\cal F})$. The following theorem gives the lower bound for this minimax risk in the inhomogeneous Besov class when dealing with multisample datasets ($i.e.$ $N>1$).

\begin{theorem}
\label{ThmBorneInf}
Under the model (\ref{FunctionalModel}) with finite variances for the processes  $E_i(\cdot)$, for $i=1,\ldots,N,$ assume that $\mu$ belongs to a Besov class ${\cal F}(s,p,q,L)$ with $p \geq 1$, $q \geq 1$, $s \geq 1/p$ and $L < \infty$, then
$$ \mathcal{R}_{N,M} ({\cal F}) \geq {\mathcal{O}\left[ \left({MN}\right)^{\frac{-s}{2s+1}} \right]} + { \mathcal{O}  \left[{M} ^{-s'} \right]}.$$
where $s' = s-1/p+1/2 > 0,$ if $p<2,$ $s' = s$ otherwise. 
\end{theorem}

Let us {mention} that the term ${(MN)}^{\frac{-s}{2s+1}}$ could be expected since it is the minimax rate ({when} $N=1$) {considering} a noise of variance  $\text{Var}(E_i(t_j))/N$. However the approximation error term ${M} ^{-s'}$, present in the case with only one sample ($N=1$), is always negligible compared with the term ${(MN)}^{\frac{-s}{2s+1}}$. When $N>1$, even a large $N$ does not provide more information on the function $\mu$ outside the grid $(t_1,\ldots,t_M)$. Hence, ${M} ^{-s'}$ becomes a limiting term.

\subsection{Wavelet estimator of the functional effect}\label{ssWE}

The upper bound of the minimax rate given in Theorem \ref{ThmBorneInf} is derived by constructing a wavelet estimator $\widehat{\mu}_{N,M}$ of $\mu$. Owing to their strong connection with the class of Besov spaces, wavelets indeed represent a powerful tool to perform adaptive functional regression (see \cite{DJK95}). 

As a brief recall and to set notations, wavelets can be used to construct orthonormal basis of the functional Hilbert space $L^2([0,1])$ by dilating and translating a compactly supported scaling function denoted by $\phi$ and a compactly supported mother wavelet denoted by $\psi$. We assume that $\phi$ and $\psi$ belongs to $C^m([0,1]).$ 
Then, {letting} $j' \in \mathbb{N}$ {be} the first level of approximation, the family:
\[
 \{ \phi_{j'k}, k=0,\ldots,2^k-1 ; \psi_{jk}, j \geq j_0, k=0,\ldots,2^k-1 \},
\]
with $\phi_{j'k}(t) = 2^{j'/2} \phi(2^{j'}t - k)$ and $\psi_{jk}(t) = 2^{j/2} \phi(2^{j}t - k)$ is an orthonormal basis of $L^2([0,1])$. Thus, any function $\mu$ in the space $L^2([0,1])$ can be expressed in the wavelet basis as:
\[
 \mu(t) = \sum_{k = 0}^{2^{j'}-1} \alpha^*_{j'k} \phi_{j'k}(t) +  \sum_{j\geq j'} \sum_{k = 0}^{2^j-1} \beta^*_{jk} \psi_{jk}(t),
\]
where $\alpha^*_{j'k} = \langle \mu,\phi_{j'k} \rangle$ and $\beta^*_{jk} = \langle \mu,\psi_{jk} \rangle$ are respectively the \emph{theorical} approximation and wavelet coefficients, and with $\langle \cdot , \cdot \rangle$ being the canonical Hilbertian scalar product associated with the space $L^2([0,1])$. In the following, we set $j'=0$ and omit the index $(0,0)$ for the unique remaining scaling coefficient denoted by $\alpha^*$. 

The Besov class ${\cal F}(s,p,q,L)$ is defined via wavelet coefficients in the following way:
$${\cal F}(s,p,q,L)=\left\{\mu\in L^2([0,1]):\; \|\mu \|_{spq} \leq L \right\},$$
where 
$$\|\mu \|_{spq}=|\alpha^*|+\left(\sum_{j=0}^\infty (2^{j(s-1/p+1/2)}\|\beta^*_{j\cdot}\|_p  \right)^\frac 1 q,\quad  \|\beta^*_{j\cdot}\|_p=\left(\sum_{k = 0}^{2^j-1} (\beta^*_{jk})^p\right)^\frac 1 p.$$
For $p,$ $q>0$ and $1/p -1<s<m,$ the norm $\|\cdot \|_{spq}$
is equivalent to the norm of the corresponding Besov space  (cf. \cite{D94}, \cite{DJ97}).

In statistical settings, we are {more} concerned with discretely sampled {curves}. 
By applying the fast discrete wavelet transform proposed by \cite{Ma89} to the functional model (\ref{FunctionalModel}), we obtain a representation of the model in the coefficient domain given by:
\begin{align}
 M^{-\frac 1 2}\Wbf \Ybf_i & = M^{-\frac 1 2}\Wbf \mubf + M^{-\frac 1 2}\Ebf_i,  \qquad \forall i=1,\ldots, N \notag \\
 \begin{bmatrix} c_i \\ \dbf_i \end{bmatrix} & = \begin{bmatrix} \alpha \\ \betabf \end{bmatrix} + \begin{bmatrix} \varepsilon^c_i \\ \varepsilonbf^d_i \end{bmatrix} \qquad \text{with } \begin{bmatrix} \varepsilon^c_i \\ \varepsilonbf^d_i \end{bmatrix} \sim \N( \zero , \Gbf). \label{WaveletModel} 
\end{align}
The $M \times 1$ vector $(c_i, \dbf_i^T)^T$ contains $empirical$ scaling and wavelet coefficients associated with the signal, while $(\alpha, \betabf^T)^T$ stand for empirical coefficients related to the fixed effect $\mu$ and $(\varepsilon^c_i, \varepsilonbf^{d \; T}_i)^T$ for the coefficients coming from the error term $\Ebf_i$.
Following \cite{AS07},  {the} modeling of  {such} correlated noise is performed directly in the wavelet domain by assuming first that $\Gbf$ is a diagonal matrix thanks to the well known decorrelating property of wavelets (see \cite{FJW92}). Then, to attain a wide range of processes, variances are assumed to vary with respect to the position and the resolution level such that $Var(\varepsilon^c_{i}) = \sigma^2_c/\sqrt{M}$ and $Var(\varepsilonbf^d_{ijk}) = \sigma^2_{jk}/\sqrt{M}$ for all $(j,k)$ in $\Lambda$ with $\Lambda = \{(j,k) | j=0, \ldots , J-1 ; k=0,\ldots, 2^j-1 \}$. Conversely, existing works dealing with a correlated noise focused on the modeling of individual noise processes in the time domain by assuming a stationnary noise (\cite{JS97}) or a locally stationnary noise (\cite{SG00}). In the wavelet domain these assumptions translate into variance terms for the matrix $\Gbf$ that are respectively depending on $j$ ($\sigma^2_{j}$) or depending on both $j$ and $k$. Based on the decorrelating property of wavelets, extra diagonal terms in the matrix $\Gbf$ are {then} neglected which restricts the class of reached processes in a way that is not effectively controlled.
As a matter of fact, our model allows to consider non stationary processes whose covariance is diagonalizable by the DWT. However, we claim that such a modeling enables to catch a wide range of processes, even non stationary and hence allows a flexible enough modeling.

In the context of inhomogeneous spaces of functions such as Besov classes, it is known that in some cases, no linear method can achieve the optimal rate (see $e.g$ \cite{HKPT98}) {whereas} nonlinear wavelet thresholding, pioneeringly introduced by \cite{DJ94} in the white noise model, achieves this goal for a wide class of functional classes by taking advantage of the natural spatial adaptivity of wavelets. 
Starting from model (\ref{WaveletModel}) in the coefficient domain, we extend the usual thresholding procedures to the heteroscedastic framework by including position-dependent variance parameters in the thresholding expressions. For $N=1,$ the wavelet coefficients $d_{1jk}$ are shrunk  {as} from a certain level, through a defined shrinkage function $\delta$, {such that} $\widehat{\beta}_{jk}=\delta(d_{1jk},\lambda_{jk}),$ where $\lambda_{jk}=\lambda\widehat{\sigma}_{jk},$ and $\lambda$ is a regularization parameter to be fixed. The shrunk coefficients are inversely transformed to yield the solution in the time domain,  {namely} $M^{\frac 1 2}\Wbf^T[\widehat{\alpha},\widehat{\betabf}^T]^T,$ where $\Wbf^T$ is the transpose of the orthogonal matrix  $\Wbf$.

When $N>1,$ \cite{AS05} propose three strategies  {to} estimate $\mu$ in model (\ref{FunctionalModel}) in the homoscedastic case. 

\begin{enumerate}
\item The most natural one, widely used in practice, is the direct pointwise averag{ing} of observations $\Ybf_1,\ldots,\Ybf_N.$ However this simple procedure leads to poor convergence rate as pointed by \cite{AS05}, reflected by the completely pointwise procedure and poor finite sample performance. This approach is referred as a simple pointwise \emph{average} approach by the authors.

\item The second approach consists in averaging the nonparametric regression curves of the $N$ signals and is referred as a \emph{shrink then average} approach. This procedure improves the convergence rate due to the presence of a smoothing step.

\item The former strategy can be further improved by first averaging the observations $\Ybf_1,\ldots,\Ybf_N$ and then apply shrinkage to the average signal using then the whole sample. That is the third approach proposed in \cite{AS05} and referred as a \emph{average then shrink} approach. Let us note that it has not been demonstrated that such an estimator achieves the optimal convergence rate. 
\end{enumerate}

In this work we consider this third approach in the heteroscedastic case and show that the associated estimator is near-minimax. Precisely, we consider
\begin{equation}
\label{ThFixedEffect}
  \widehat{\mubf}_{N,M} = M^{\frac 1 2} \Wbf \begin{bmatrix} \widehat{\alpha} \\ \widehat{\betabf} \end{bmatrix}, 
\end{equation}
{with},
\begin{align}
 \widehat{\alpha} & =  c_{\bullet},  \label{coeff} \\
 \widehat{\betabf} & = \left \lbrace
	\begin{array}{ll}
 d_{\bullet,jk},  &  \text{for} \; \; j < j_0, \\
  \delta (d_{\bullet jk}, \lambda_{jk}), &  \text{for} \; \; j=j_0, \ldots,J-1, \notag 
	\end{array}
\right.
\end{align}
where $\bullet$ denotes the average over the $N$ samples.
The choice of the parameter $j_0$ will be detailed in the proof of Theorem \ref{ThmBorneSup}. The values of position-dependent thresholds $\lambda_{jk}$ are then given by:
\begin{equation}
\label{lambdajk}
 \lambda_{jk} = \widehat{\sigma}_{jk} \frac{\sqrt{2 \log M}}{\sqrt{M}},
\end{equation}
where $\widehat{\sigma}_{jk}^2$ are $\sqrt{N}$-consistent estimates of variances.
The following result gives an upper bound for the quadratic risk depending on the signal size $M$ and the number of samples $N$.

\subsection{Upper bound of the minimax risk for wavelet-based thresholding estimators}\label{ssUB}

\begin{theorem}
\label{ThmBorneSup}
Under the model (\ref{FunctionalModel}), assume that $\mu$ belongs to a Besov class ${\cal F}(s,p,q,L)$ with $p \geq 1$, $q \geq 1$, $s \geq 1/p$, $L < \infty,$ and that the variances $\sigma^2_c$ and $(\sigma^2_{jk})_{(j,k) \in \Lambda}$ are bounded by a constant denoted by $\sigma^2_{max}$. For any shrinkage function that satisfies, for any $\beta$ and $\xi,$
\begin{equation}
\label{contdelta}
|\delta(\beta+\xi,\lambda)- \beta | < C(\min \left(|\beta|,\lambda)+ |\xi|\mathbf{1}_{|\xi|>\lambda/2}\right), 
\end{equation}
then the estimator $\widehat{\mu}_{N,M}$ defined by (\ref{ThFixedEffect},\ref{coeff},\ref{lambdajk}) with  thresholds $2\lambda_{jk}$, satisfies
$$ \E \big( \| \widehat{\mu}_{N,M} - \mu \|_{2}\big) \leq 
\left\{
\begin{aligned}
 & \max \left\{ \mathcal{O}\left[ \left(\frac{\log M}{MN}\right)^{\frac{s}{2s+1}} \right] + \left[\mathcal{O} \left( \frac{\log M}{M} \right)^{s'} \right] \right\}, \; \text{if} \; \; \frac{2}{2s+1} < p < 2\\
 & \max \left\{ \mathcal{O}\left[ \left(\frac{1}{MN}\right)^{\frac{s}{2s+1}} \right] + \left[\mathcal{O} \left( \frac{\log M}{M} \right)^{s'} \right] \right\},  \; \text{if} \; \; p \geq 2
\end{aligned}
\right.
$$
where $s'$ is  defined as in Theorem \ref{ThmBorneInf}.
\end{theorem}

The next section describes the practical derivation of thresholding procedures that satisfy the conditions required by Theorem \ref{ThmBorneSup}. Thus we propose estimators that enjoy a near-optimal convergence rate in a multisample heteroscedastic setting.

\section{Thresholding strategies}
\label{ThStrategies}

\subsection{Shrinkage functions}
Among the usual thresholding procedures we first focus on the hard and soft thresholding procedures of \cite{DJ94}, that provide estimators $\widehat{\beta}^{\text{h}}$ and $\widehat{\beta}^{\text{s}}$. We also consider the SCAD ($\widehat{\beta}^{\text{scad}}$) thresholding of \cite{AF01} that establishes a trade-off between hard and soft thresholding, overcoming their respective {non}-continuity and bias drawbacks. The main conclusion of Theorem \ref{ThmBorneSup} is subject to the fulfilling of constraint (\ref{contdelta}). The Lemma 2 of \cite{JD96} ensures that hard and soft thresholding meet this requirement. Moreover, since we have
\[
\forall \beta \in \rset, \forall \lambda >0, \;\;  \delta^{\text{s}} (\beta, \lambda) \leq \delta^{\text{scad}} (\beta, \lambda) \leq \delta^{\text{h}} (\beta, \lambda),
\]
the conclusion of this Lemma still holds for the SCAD thresholding.

\subsection{Choice of the threshold}

For theoritical purposes, only the universal threshold (\ref{lambdajk}) has been considered so far. Its easy implementation and its good asymptotic properties makes the universal threshold very popular in major wavelet packages. Our heteroscedastic thresholding approach is based on the definition of a threshold depending on the position $(j,k)$ through the variance parameters (see (\ref{lambdajk})). Theorem \ref{ThmBorneSup} then applies in the context where the variances are unknown but for which $\sqrt{N}$-consistent estimates are available. 
When $N=1$, exhibiting $\sqrt{N}$-consistent variance estimates is challenging. Such an issue has been considered in the litterature and approaches based on a functional modeling of the variances in the time domain have been developed (see $e.g.$ \cite{GSJ89}, \cite{AL95}, \cite{CW08}). In their approaches, variances are then estimated using $\nu$-order differences ($\nu \in \nset$), coupled with an appropriate smoothing nonparametric method.

In the mutlisample context ($N>1$), variance parameters can be easily estimated by simply considering empirical variances estimators such that:
\begin{equation}
\label{EstSigma}
  \widehat{\sigma}^2_{jk} = \frac{1}{N-1} \sum_{i=1}^{N} (d_{ijk} - d_{\bullet jk})^2, \qquad \text{for all } (j,k) \in \Lambda.
\end{equation}
These variance parameter estimates straightforwardly satisfy the $\sqrt{N}$-consistency requirement due to their asymptotic normality properties.

However as pointed by \cite{DJ94} and \cite{CoD95} the universal threshold, originally designed for a "noise-free" reconstruction, is substantially larger than the minimax threshold.
To handle this practical drawback, \cite{DJ95} proposed a strategy based on the Stein Unbiased Risk Estimate (SURE, \cite{St81}) whose purpose is to fix level dependent thresholds $ \lambda_{\text{SURE},j}$ that leads to obtain an unbiased estimate of the $L^2$-risk. 
Let $\widetilde{\dbf}$ be a vector in $\rset^\ell$ distributed as a standardized Gaussian distribution of mean $\betabf$ and covariance matrix equal to identity. The idea consists in writing the thresholding estimator $\widehat{\beta}(\cdot)=\delta(\cdot,\lambda)$ as the sum:
\[
\widehat{\beta}(\widetilde{\dbf}) = \widetilde{\dbf} + \gbf(\widetilde{\dbf}),
\]
where $\gbf$ is a weakly differentiable function from $\rset^\ell$ to $\rset^\ell$. This leads to:
\[
 \E \left( \|  \widehat{\beta} (\widetilde{\dbf}) - \beta \|^2_{2} \right) = \ell +  \E \left( \| \gbf (\widetilde{\dbf}) \|^2_{2} + 2 \sum_{k=1}^\ell \frac{\partial \gbf (\widetilde{\dbf})}{\partial {d}_k} \right).
\]
The goal is then to select the parameter $\lambda$ which minimizes the estimate of the $L^2$-risk, denoted by SURE$(\lambda;\widetilde{\dbf})$ and given by
 \[
 SURE(\lambda;\widetilde{\dbf}) = \ell +  \| \gbf (\widetilde{\dbf}) \|^2_{2} + 2 \sum_{k=1}^\ell \frac{\partial \gbf (\widetilde{\dbf})}{\partial {d}_k}.
\]
By considering $\widetilde{d}_{jk} = d_{\bullet,jk} / \hat{\sigma}_{jk}$, where $\hat{\sigma}^2_{jk}$ is given as in (\ref{EstSigma}),
the SURE threshold is given by:
\begin{equation}
\label{SURE.Th}
   \lambda_{\text{SURE},j} = \argmin_{0 \leq \lambda \leq \lambda_{U,j}} \text{SURE} (\lambda,\widetilde{\dbf}_j) \qquad \text{for all} \; j = j_0, \ldots, J-1,  
\end{equation}
where $\lambda_{U,j}$ is the universal threshold given in (\ref{lambdajk}) and $M=2^j.$
The computation of the SURE criterion depends on the chosen thresholding function. Following the example of \cite{DJ95} for soft thresholding, we propose an adaptation of the SURE concept to SCAD thresholding. Let us note that \cite{Pa10} proposed an other derivation of the SURE criterion leading to a SURE-Block-SCAD estimator in the context of wavelet-based functional regression. When replicates are available, the SURE criterion to minimize according to $\lambda$ is given by:
\begin{multline}
 \label{SURE.SCAD}
 \text{SURE}_{\text{SCAD}} (\lambda;\widetilde{\dbf}_j) = 2^j + \sum_{k=0}^{2^j - 1} (\widetilde{d}_{jk}^2-2) \mathbf{1}_{ \{ | \widetilde{d}_{jk}| \leq \lambda \} } +  \sum_{k=0}^{2^j - 1} \lambda^2 \mathbf{1}_{ \{ \lambda < | \widetilde{d}_{jk}| \leq 2\lambda \} } \\ + \frac{1}{(a-2)^2} \sum_{k=0}^{2^j - 1} \left[ 2(a-2) + \widetilde{d}_{jk}^2 + (a\lambda)^2 + 2a\lambda |\widetilde{d}_{jk}^2| \right] \mathbf{1}_{ \{ 2\lambda < | \widetilde{d}_{jk}| \leq \lambda \} }.
\end{multline}
The computation details can be found in Appendix \ref{appendix:SURE.SCAD}. As recommended by \cite{FL01}, $a$ is set to $3.7$ based on a Bayesian argument. 

Moreover, we can point out that extremely sparse settings can lead to insufficient denoising due to the impact of zero coefficients in SURE criterion. To avoid this drawback, \cite{DJ95} propose a compromising Hybrid Scheme (HS) between regular and SURE thresholding defined by:
 \begin{equation}
  \label{SURE.SCAD.HS}
  \lambda_j^\text{HS} = 
  \left\{
  \begin{array}{ll}
   \lambda_{U,j} \; & \text{if} \; \sum_{k=0}^{2^j - 1} d_{\bullet,jk}^2 \leq \widehat{\sigma}^2_{jk} 2^{j/2}(2^{j/2} + j^{3/2}), \\
   \lambda_{\text{SURE},j} \; & \text{otherwise},
  \end{array}
  \right.
  \end{equation}
  for all $j = j_0, \ldots, J-1$. 

\section{Numerical experiments}
\label{Applications}

In this simulation study, we first investigate the benefits of using heteroscedastic thresholding estimators over homoscedastic ones when more than one sample are available. Then, we investigate the effect of the choice of the threshold on realistic simulated datasets.

\paragraph{Simulation settings.} We consider the test functions \texttt{Blocks}, \texttt{Bumps}, \texttt{Heavisine} and \texttt{Doppler} \citep{DJ94} that we use to model the principal mean functions $\mu$. These functions are processed with the Daubechies' extremal phase wavelet basis with respectively 1,2,5 and 7 vanishing moments, based on the Shannon entropy as described in \cite{Na08}, Chap 2. Then we get the noise-free wavelet coefficients, to which we add heteroscedastic noise, following model (\ref{WaveletModel}): multisamples are simulated in the wavelet domain by corrupting the wavelet coefficients of the mean function by a normally additive heteroscedastic noise whose variance $\sigma^2_{jk}$ at a given position $(j,k)$ in $\Lambda$ is given by:
\begin{equation}
\label{HeteroStruct}
\sigma^2_{jk} = 
\left\{
\begin{aligned}
& \sigma^2  & \quad \text{for} \; \; (j,k) \in \Lambda_1,\\
& \sigma^2 + 2^{-j\eta}\gamma^2_{jk} & \quad \text{for} \; \; (j,k) \in \Lambda_0.
\end{aligned}
\right.
\end{equation}
The set $\Lambda_0 \subset \Lambda$ contains index associated with the zero coefficients of the mean function whereas $\Lambda_1$ contains the ones associated with nonzero coefficients. The first term $\sigma^2$ is associated to a white noise added to all coefficients, whereas the second term is an extra variability that introduces heteroscedasticity at some positions. Following \cite{AS07}, a scale-wise exponential decrease is imposed to the extra variability terms by the quantity $2^{-j\eta}$. Parameter $\eta$ relates to the fixed effect regularity allowing the extra variability associated to $\gamma^2_{jk}$ to remain interpretable. In the following we use $\eta = 1.5$. \\


\paragraph{Dealing with zero and non-zero coefficients.} One expects heteroscedastic thresholding estimators to be favored by heteroscedasticity structure expressed on the zero coefficients of the mean function: true zero coefficients are indeed more susceptible to be thresholded in this setting since heteroscedastic thresholds are expected to be larger than the homoscedastic one. Therefore we put emphasis on configurations where heteroscedasticity concerns the null wavelet coefficients of the mean function.

The value of $\sigma^2$ is controlled by a Signal-to-Noise Ratio (SNR) and takes values in (1,5) going from a high level (SNR=1) to a low level of noise (SNR=5). Parameters $\gamma^2_{jk}$ are then drawn from a Gamma distribution with scale 2 and shape $\gamma^2_\text{ref}/2$. The quantity $\gamma^2_\text{ref}$ associated to the heteroscedasticity intensity is controlled with respect to the baseline variance $\sigma^2$ by a ratio parameter $\tau$ defined by
\[
 \tau = \frac{M \sigma^2}{\gamma^2_\text{ref} \sum_{(j,k)\in \Lambda_0}  2^{-j\eta}}.
\]

\paragraph{Parameters values.} We set the signal size to $M=2048$ and the sample size to $N=100$. A wider simulation study (not shown for the sake of clarity) reveals that the main conclusions do not differ with different signal and sample size. For each fixed effect function, the simulation design explores the following configurations: SNR $ \in (1,5)$, $\tau \in (0.1,1)$. The variability and heteroscedasticity parameters $\sigma^2$ and $\gamma^2_\text{ref}$ are deduced from the value of SNR and $\tau$ respectively. Each configuration is repeated 200 times.

\paragraph{Heteroscedastic versus homoscedatic thresholding.} We start by considering the framework defined by the assumptions of Theorem \ref{ThmBorneSup}, $i.e.$ we consider the SCAD thresholding function with the universal threshold in a heteroscedastic setting. Since the threshold used in Theorem \ref{ThmBorneSup} is known to be large \citep{DJ94}, it is set to half of its value in the following. Then heteroscedatic thresholding (denoted He) refers to the procedure that uses empirical estimates of the variance at each position $(j,k) \in \Lambda$ whereas homoscedastic thresholding (denoted Ho) uses $\widehat{\sigma}^2_{MAD}$ (based on the Median Absolute Deviation (MAD) of the coefficients at the finest resolution level \citep{DJ94}). \cite{AS05} introduced the idea of wavelet-based thresholding in the context of noisy repeated measurements and discussed how to integrate the replicates in the analysis. They use in (\ref{lambdajk}) the usual robust variance estimate $\widehat{\sigma}^2_{MAD}$ instead of  position-dependent estimators $\widehat{\sigma}^2_{j,k}.$ However, they do not investigate the effect of the choice of the threshold, and they do not handle the potential heteroscedasticity in their synthetic data, despite the presence of inter-individual variability. A simulation study (not shown) revealed that the strategy of taking the mean of the individual MAD leads to better performance. Therefore we consider this strategy for the homoscedastic part. We aim at comparing homoscedastic and heteroscedastic procedures regarding to the mean function reconstruction performance. Performance of competed procedures are evaluated with respect to the Mean Integrated Squared Error (MISE) of the reconstructed mean function.

\subsection{Results.}

Average MISEs are presented on Table \ref{Table:ExResHomoVsHetero}. The results show that heteroscedatic estimates greatly outperform homoscedastic ones in terms of functional reconstruction for all considered configurations. As expected, this is especially true when the heteroscedasticity intensity is high ($i.e$ for $\tau=0.1$). 

\begin{center}
Table 1 here
\end{center}


Another argument supporting the use of heteroscedastic thresholding procedures concerns their adaptative behaviour in an homoscedastic framework: indeed, a simulation study in the homoscedastic framework ($i.e.$ with $\sigma^2_{jk} = \sigma^2$ for all $(j,k) \in \Lambda$) reveals similar reconstruction properties of homoscedastic and heteroscedastic estimates for a SCAD thresholding using the universal threshold.
Corresponding results are displayed in Table~\ref{Table:ExResHomo}.

\begin{center}
Table 2 here
\end{center}

\paragraph{Comparing heteroscedastic procedures}

Despite good asymptotic properties, using the universal threshold may not be optimal in finite dimensional setting as mentioned by \cite{DJ94} in their original paper. Therefore we now focus on comparing heteroscedastic procedures for different choices of thresholds on simulated datasets. In order to consider more realistic cases, we consider datasets where heteroscedasticity corrupts both null and non null coefficients of the mean function. Hence, starting from the same mean functions, the heteroscedasticity is as from now defined such that for $(j,k)$ in $\Lambda$:
\begin{equation}
\label{VarSimu}
 \sigma^2_{jk} = \sigma^2 + \pi_{jk} \times 2^{-j\eta} \gamma^2_{jk}.
\end{equation}
The quantities $\sigma^2_{jk}$ and $\gamma^2_{jk}$ are defined as previously whereas $\pi_{jk}$ is assumed to be a realization of a Bernoulli distribution with parameters 0.3. Note that the pairs fixed effects-$\mu$/heteroscedasticity structure-$\pibf = (\pi_{jk})_{(j,k) \in \Lambda}$ are kept fixed for all the synthetic datasets.

\begin{figure}[ht!]
\begin{center}
\includegraphics[scale=0.45,angle=-90]{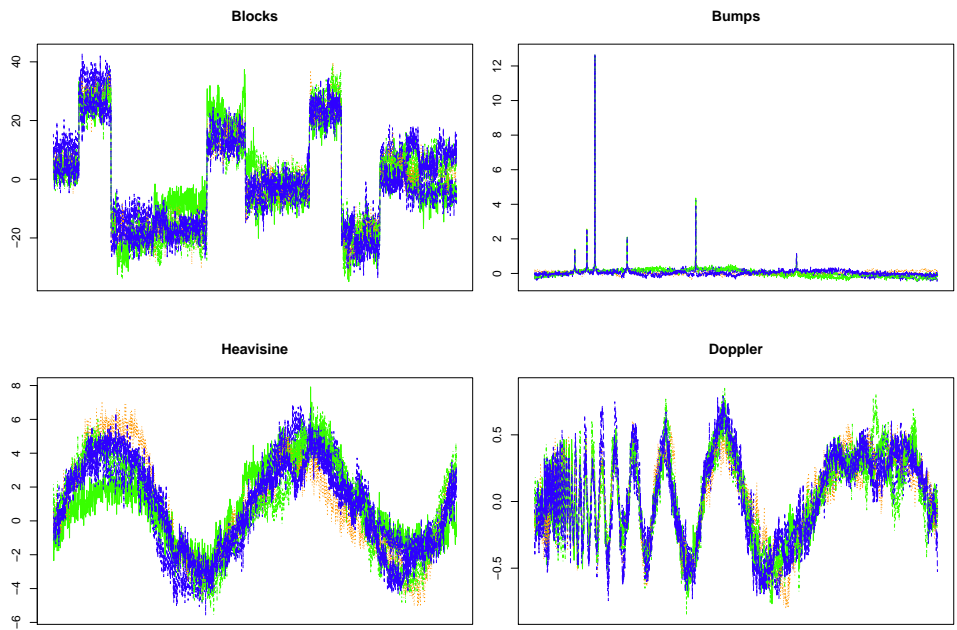}
\end{center}
\caption{Examples of realistic simulated data. For each mean functions \texttt{Blocks}, \texttt{Bumps}, \texttt{Heavisine} and \texttt{Doppler}, 5 random realizations are represented. The parameters SNR and $\tau$ are respectively set to 5 and 0.25. \label{Fig:ExSimData}}
\end{figure}

For each mean function associated to a given heteroscedastic structure $\pibf$, the simulation design explores the following configurations: SNR varies in $(1,5)$ and $\tau$ in $(0.1,0.25,1)$. Similarly, the signal and sample size are respectively set to $M=2048$ and $N=100$ whereas each configuration is repeated 200 times. 
Examples of simulated data are represented on Figure \ref{Fig:ExSimData} for all considered main patterns.

\begin{center}
Figure 1 here
\end{center}

\begin{figure}[ht!]
\begin{center}
	\includegraphics[scale=0.6]{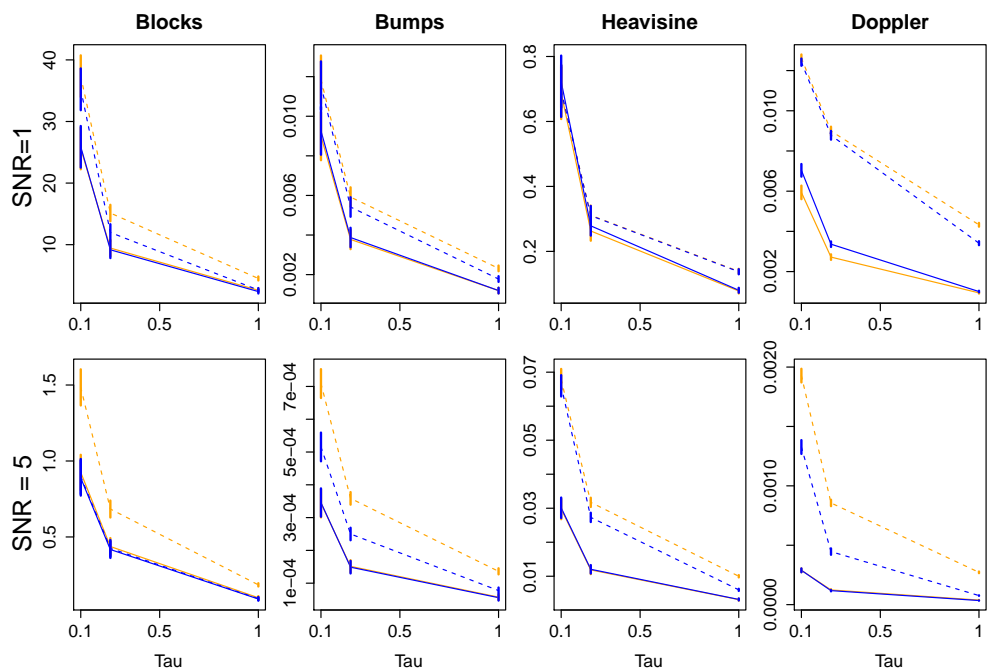}
\end{center}
\caption{Resulting MISEs averaged over 200 repetitions for reconstructed fixed effects. Two SNR values in rows (SNR = (1,5) for a high/low noise) and three heteroscedasticity intensities on the horizontal axis of each graph ($\tau$ = 0.1,0.25,1 from a high level to a low level) are considered. Soft and SCAD thresholding functions differ by plotting colors (respectively in orange and blue) whereas threshold choices  Universal and SURE differ by the line types (respectively in dashed and solid line). Vertical bars are associated to the standard deviations of the resulting MISEs. \label{Fig:ResMISEs}}
\end{figure}

\begin{figure}[ht!]
\begin{center}
\includegraphics[scale=0.5,angle=-90]{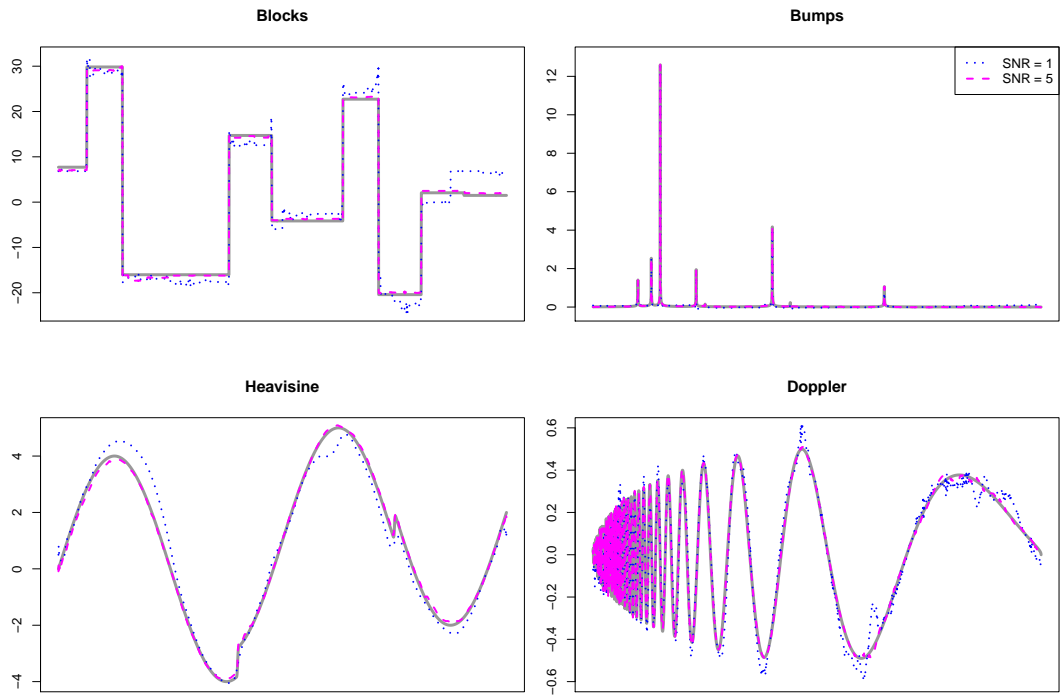}
\end{center}
\caption{Examples of reconstructed mean functional effect using an heteroscedastic SCAD thresholding with the SURE threshold for models \texttt{Blocks}, \texttt{Bumps}, \texttt{Heavisine} and \texttt{Doppler}. The true mean functions is displayed in plain gray line. The parameter $\tau$ is equal to 0.25 whereas SNR take the values 1 (for a high noise, displayed in dotted blue lines) and 5 (for a low noise, displayed in dashed magenta lines). In all configurations, the chosen realization correspond to the one giving rise to the median MISE.\label{Fig:ExReconst}}
\end{figure}

Then heteroscedastic thresholding procedures are competed for both Soft and SCAD thresholding functions, $\delta^{\text{s}}$ and $\delta^{\text{scad}}$, and for both Universal and SURE threshold, $\lambda_U$ and $\lambda_{HS}$. Performance of the procedures are evaluated with respect to the Mean Integrated Squared Error (MISE) of the reconstructed mean functions.
Simulation results are presented on Figure~\ref{Fig:ResMISEs}. Examples of reconstruction associated to median performance are represented in Figure \ref{Fig:ExReconst} 

\begin{center}
Figure 2 and Figure 3 here
\end{center}


As a main conclusion we can observe that using the SURE threshold leads to improved performance for the reconstruction of the main effect in a heteroscedastic setting. As mentionned by \cite{DJ95} in the homoscedastic framework, the universal threshold turns out to be too large in practice when dealing with finite dimensional signals. 

Another interesting point concerns the interaction between the choice of the threshold and the thresholding function. When using the universal threshold, the SCAD thresholding gives indeed at least similar or improved reconstruction performance. This is expected since the SCAD thresholding is designed to smoothly correct the bias on high coefficients introduced by the soft thresholding. 
Conversely such a difference vanishes when using the SURE threshold for which Soft and SCAD thresholdings exhibit similar performance. This finding can be explained by the adaptative behaviour of the SURE threshold that compensates the existing bias on high coefficients.

By way of conclusion, the overall simulation study encourages the use of the heteroscedatic thresholding in the context of functional regression with multiple samples. Heteroscedastic thresholding keeps indeed the simplicity and the computational efficiency of the usual homoscedastic thresholding while being able to handle potential inter-individual variations. Moreover, in practice, using the adaptative SURE threshold, paired with the SCAD thresholding which enjoys good theoritical properties leads to improved reconstruction of the mean function.  

As a last remark, we shall mention that the wider simulation study abovementioned with various sample and signal sizes shows that the overall MISEs orders of magnitude are more improved by a higher number of samples $N$ than by a larger signal size $M$.

\subsection{Analysis of experimental data}

As an application to the proposed methodology, we analysed a SELDI-TOF mass spectrometry dataset issued from a study on ovarian cancer \citep{PAH02}. This dataset was produced by the Ciphergen WCX2 protein chip and is publicly available through the Clinical Proteomics Programs Databank (\footnote{\url{http://home.ccr.cancer.gov/ ncifdaproteomics/ppatterns.asp}}, ovarian dataset 8-7-02). The sample set consists of 162 serums profiles from women affected by an ovarian cancer and 91 control subjects. Each spectra contains the measure of 15154 intensities characterizing as many mass over charge ($m/z$) ratios. Prior to analysis, raw data are background corrected using a quantile regression procedure, and spectra are aligned using a procedure based on wavelets zero crossings \citep{ABL07}. Moreover, we restrict on 512 intensities for $m/z$ ratios within the range [5200,5915] centered around the main central peak.
Mass spectrometry data represented a meaningful application for our method since \cite{GLM13} show evidence for the presence of inter-individual variations occuring at specific ranges of $m/z$ ratios resulting in a sharp heterosecdasticity structure. 

\begin{figure}[ht!]
\begin{center}
\includegraphics[width=10cm,angle=-90]{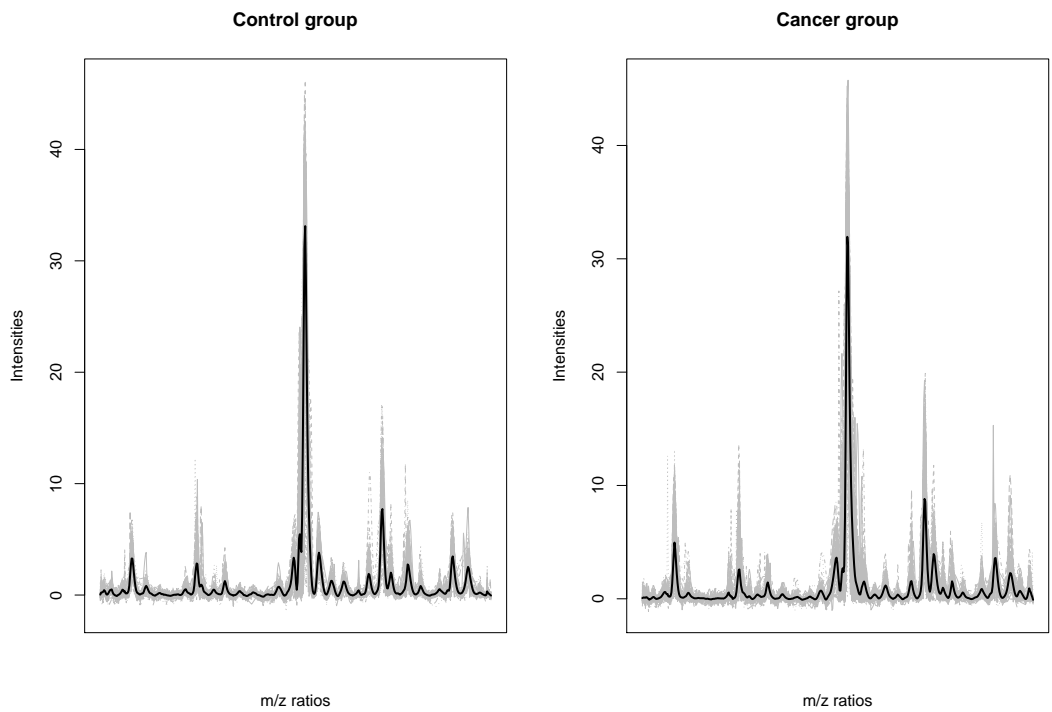}
\end{center}
\caption{Mean reconstructed functions (bold line) superimposed on observed data (in light gray) for the control group and the cancer group.  \label{Fig:MeanSpectro}}
\end{figure}

We separately analysed the control group and the group affected by a cancer using an heteroscedastic SCAD thresholding procedure, with a SURE threshold. Mean reconstructed functions superimposed on experimental data are represented in Figure \ref{Fig:MeanSpectro}. 

\begin{center}
Figure 4 here
\end{center}

We can observe that individuals from the control and cancer groups exhibit similar mean functional profiles. Such an observation indicates that a nonparametric testing procedure would be on purpose to ascertain the presence of a significant effect of the group. Although it is out of the scope of the present paper, in this context, taking into account the presence of potential inter-individual variations appears as critical for the application of such testing procedure.

\section*{Acknowledgements}
Part of this work was supported by the Interuniversity Attraction
Pole (IAP) research network in Statistics P5/24. We are grateful to
Anatoli Juditsky for constructive and fruitful discussions.
\begin{comment}
\end{comment}

\bibliographystyle{chicago}

\section{Appendix}

\subsection{Proof of Theorem \ref{ThmBorneInf}}

First let us recall the aim of the proof {concerning the} lower bound in the minimax context. Since
$$ \mathbb{E} \big( \mathcal{R}_{N,M}^{-1} ({\cal F})\| \widehat{\mu}_{N,M}-\mu \|_2 \big) \geq c_1\mathbb{P}\left( \| \widehat{\mu}_{N,M}-\mu \|_2\geq c_1\mathcal{R}_{N,M}({\cal F})\right),$$ 
for some $c_1>0,$ we have to show that
$$ \mathbb{P}\left( \| \widehat{\mu}_{N,M}-\mu \|_2\geq c_1\mathcal{R}_{N,M}({\cal F})\right)>c_2,$$
for some constant $c_2>0$. Next we reduce the class ${\cal F}$ to a subclass  ${\cal F}_n$ of finite number $n$ of functions in ${\cal F}$ because the $sup$ is greater over a larger class. The family ${\cal F}_n= \{\mu_0, \ldots, \mu_{n-1} \}$ is constructed by small perturbation of $\mu_0$, so that {the distance between each pairs of functions is small and at least of order $\mathcal{R}_{N,M}({\cal F}).$}
 Then the problem can be reduced to {the one} of testing by the following way
$$\sup_{\mu \in {\cal F}_n} \mathbb{P}\left( \| \widehat{\mu}_{N,M}-\mu \|_2\geq c_1\mathcal{R}_{N,M}({\cal F})\right)\geq p_n= \inf_{\phi}\max_{j=0,\ldots, n-1} {\pi_{\phi} (\mu=\mu_j)},$$
{with $\pi_{\phi}$ the power function associated to $\phi$}, where $\phi$ is any test that allows to distinguishing between the $n$ hypotheses, {the} $k$-th of them stating that the observations of model (\ref{FunctionalModel}) are drawn
from the $k$-th element of the set ${\cal F}_n$. To bound $p_n$ by $c_2>0,$ we need to major the maximum of the Kullback distance $K(\mu_i,\mu_j)$ between observations of model (\ref{FunctionalModel}) associated with $\mu_i$ and the ones associated with $\mu_j$. For instance when $n=2,$ we have
$$p_2\geq \max \left(\frac {\exp(-K(\mu_1,\mu_0))}4, \frac{1-\sqrt\frac {K(\mu_1,\mu_0)}2}{2}\right).$$
Without loss of generality, {since variances are assumed to be bounded}, we can consider model (\ref{FunctionalModel}) with $E_i(t_j)$, $i=1,\ldots, N,$ $j=1,\ldots, M,$ independent and identically distributed Gaussian random variables  with zero-mean and variance $\sigma^2_{E}.$ In this case, we have
\begin{equation}
  \label{Kulb}
K(\mu_1,\mu_0)=\frac N {2\sigma^2_{E}} \sum_{j=1}^M (\mu_1(t_j)-\mu_0(t_j))^2.
\end{equation}
Let us come back to the proof of the lower bound. This proof can be decomposed in two steps. For the usual term in  $\mathcal{O}\left[ \left({MN}\right)^{\frac{-s}{2s+1}} \right],$ we {just have} to use the usual proof for the Besov classes by adding the factor $N$ because of the multiplicative term $N$ in (\ref{Kulb}).  We now give the proof corresponding to the term in $\mathcal{O}\left[ {M}^{{-s'}} \right].$ We {only need} two functions in order to construct ${\cal F}_n,$ that is $n=2.$ 
For $p\geq 2,$ we put $\mu_0(t)=0,$ for all $t\in [0,1],$ and
$$\mu_1(t)= {M}^{\frac 1 p -s}\eta\left(Mt-\frac 12 \right),$$
where $\eta\in {\cal F}(s,p,q,L)$ with support equal to $[-1/2,1/2]$ such that $\eta(-1/2)=\eta(1/2)=0,$ and $\|\eta\|_2\geq c>0$. We have $\mu_0\in {\cal F}(s,p,q,L)$ and
$$\| \mu_1 \|_{spq} \leq {M}^{\frac 1 p -s}  M^{s-\frac 1 p} \| \eta \|_{spq}\leq L.$$
So we {also have} $\mu_1\in {\cal F}(s,p,q,L),$ and
$$\| \mu_1-\mu_0 \|_2=M^{-\frac 1 2}{M}^{\frac 1 p -s} \|\eta\|_2\geq {M}^{-s'}c,$$
{hence}, the family ${\cal F}_2$ is inclued in the Besov class ${\cal F}(s,p,q,L)$ and the $L_2$-distance between the two functions are at at least $M^{-s'}.$ Since 
$$K(\mu_1,\mu_0)=\frac N {2\sigma^2_{E}} {M}^{\frac 2 p-2s}\eta^2\left(M t_1-\frac 12 \right)=\frac N {2\sigma^2_{E}} {M}^{\frac 2 p-2s}\eta^2\left(\frac 1 2 \right)=0,$$
we have $$p_2\geq 1/2,$$ and 
$$ \mathbb{E} \big(\| \widehat{\mu}_{N,M}-\mu \|_2 \big) \geq \frac c 2 M^{-s'},$$
for any estimator $\widehat{\mu}_{N,M}.$

For $p\geq 2,$ we use the same {method} but by choosing
$$\mu_1(t)= {M}^{-s}\sum_{j=1}^M\eta\left(Mt-j+\frac 12 \right).$$
So we have
$$\| \mu_1 \|_{spq} \leq {M}^{-s} M^{\frac 1 p} M^{s-\frac 1 p} \| \eta \|_{spq} \leq L,$$
$$\| \mu_1-\mu_0 \|_2=M^{\frac 1 2}M^{-\frac 1 2}{M}^{-s} \|\eta\|_2\geq {M}^{-s}c,$$
and
$$K(\mu_1,\mu_0)=\frac N {2\sigma^2_{E}} {M}^{-2s}\sum_{j=1}^M\sum_{k=1}^M \eta\left(M t_j-k+\frac 12 \right)=\frac N {2\sigma^2_{E}} {M}^{-2s+1}\eta\left(\frac 1 2 \right)=0,$$
{which} concludes the proof.

\subsection{Proof of Theorem \ref{ThmBorneSup}}

This proof is an adaptation of the proof of Theorem 1 of \cite{JD96}.
We denote by $\widetilde{\beta}_{jk}$, the estimator $\widehat{\beta}_{jk}$ with $\sigma_{jk}$ instead of $\widehat{\sigma}_{jk}$ in(\ref{lambdajk}) and $\widetilde{\mu}_{N,M}$ the associated estimator. We introduce
$$\mu_{j_1}(t) = \alpha\phi_{00}(t) +  \sum_{j=0}^{j_1} \sum_{k = 0}^{2^j-1} \beta_{jk} \psi_{jk}(t),$$
where $j_1$ is such that $M / \log M \leq 2^{j_1} \leq 2M / \log M.$ Let us note that (see proposition 1 of \cite{DJ97}), there exists some constant $C_0$ such that this function belongs to ${\cal F}(s,p,q,C_0L).$
The global quadratic risk can then be decomposed such that:
\begin{align}
 \E \left( \| \widehat{\mu}_{N,M} - \mu \|_{2}^2  \right) & \leq \E \left( \| \widehat{\mu}_{N,M} - \widetilde{\mu}_{N,M} \|_{2}^2 \right) + \E \left( \| \widetilde{\mu}_{N,M} - \mu_{j_1} \|_{2}^2 \right) +\| \mu -\mu_{j_1} \|_{2}^2  
\notag \\
 & \leq   \E \left[ \sum_{j=j_0+1}^{j_1} \sum_k | \widehat{\beta}_{jk} - \widetilde{\beta}_{jk} |^2 \right] \notag \\
 & \quad \; + \E (|\widehat{\alpha} - \alpha |^2) + \E \left[ \sum_{j=0}^{j_0} \sum_k | \widetilde{\beta}_{jk} - \beta_{jk}|^2 \right] \label{RiskL2} \\ 
  & \qquad + \E \left[ \sum_{j=j_0+1}^{j_1} \sum_k | \widetilde{\beta}_{jk} - \beta_{jk}|^2 \right] 
+\| \mu -\mu_{j_1}\|_{2}^2 \notag \\
 & = T_1 + T_2 + T_3 + T_4 + T_5. \notag
\end{align}
We seek to bound from above each term of the decomposition. 
By using the delta method based on a Taylor expansion of the thresholding function and since $\widehat{\sigma}_{jk}^2$ are $\sqrt{N}$-consistent estimates of variances, we get:
\begin{align}
  T_1 & \leq \sum_{j=j_0+1}^{j_1} \sum_k C_1 \; \frac{2N-1}{(MN)^2} \; \sigma_{jk}^4 \notag \\
  & \leq C_1 \sigma^4_{\text{max}} \; \frac{2^{j_1}}{M^2N} \leq C_1 \sigma^4_{\text{max}} \; \frac{(\log M)^{-1}}{MN}, \notag
\end{align}
with $C_1$ being a positive constant. The model (\ref{WaveletModel}) leads to
$$
c_\bullet \sim \N \left[ \alpha, \frac{\sigma^2_c}{NM} \right] \quad \text{and} \quad   d_{\bullet, jk}\sim \N \left[ \beta_{jk}, \frac{\sigma^2_{jk}}{MN} \right].
$$
Approximation coefficients in $T_2$ are {left} unchanged, hence we have:
\[
 T_2 = \E (|c_\bullet - \alpha |^2) \leq \frac{\sigma^2_{\text{max}}}{MN}.
\]
In the same way, terms in $T_3$ are not thresholded, hence we get:
$$ T_3 \; = \; \E \left[ \sum_{j=0}^{j_0} \sum_k | \widetilde{\beta}_{jk}  - \beta_{jk}|^2 \right]  =  \sum_{j=0}^{j_0} \sum_k \E \left( |  d_{\bullet jk} - \beta_{jk}|^2 \right)  \leq C_3 \; 2^{j_0} \frac{\sigma^2_{\text{max}}}{NM} , $$
with $C_3$ being a positive constant.
The term $T_5$ is the approximation term that can be bounded such that (see proposition 1 of \cite{DJ97}):
$$ T_5 \leq C_5 2^{-2j_1s'} \leq \left[ \frac{\log M}{M} \right]^{2s'}. $$
Finally, bounding term $T_4$ needs the use of constraint (\ref{contdelta}) with $\lambda'=2\lambda$, we have
\begin{multline}
T_4 \leq \quad \underbrace{\E \left[ \sum_{j=j_0+1}^{j_1} \sum_k \min \left( |\beta_{jk}|,  \frac{ \lambda' \sigma_{jk}}{\sqrt{MN}} \right)^2 \right]}_{T_{4.1}} \\ + \underbrace{\E \left[ \sum_{j=j_0+1}^{j_1} \sum_k |\varepsilon^d_{\bullet jk}|^2 \mathbf{1}_{|\varepsilon^d_{\bullet jk}|> \frac{\lambda' \sigma_{jk}}{2\sqrt{MN}}} \right]}_{T_{4.2}},\notag
\end{multline}
where $\varepsilon^d_{\bullet jk}=d_{\bullet jk}-\beta_{ jk}.$
For the term $T_{4.1}$, since $\mu_{j_1}\in{\cal F}(s,p,q,C_0L)$, we obtain:
\begin{align}
 T_{4.1} & \leq \sum_{j=j_0+1}^{j_1} \sum_k \left(2 \lambda \frac{\sigma_{\text{max}}}{\sqrt{MN}} \right)^{2-p} \; |\beta_{jk}|^p \notag \\
 & \leq C_{4.1} \left(\frac{2\log M}{M} \right)^{1-\frac{p}{2}} \left(\frac{\sigma^2_{\text{max}}}{N} \right)^{1-\frac{p}{2}}  \underbrace{\sum_{j=j_0+1}^{j_1} \sum_k |\beta_{jk}|^p}_{= \gdO (2^{-s'pj_{0}})} \notag \\
 & \leq C_{4.1} \; \left(\frac{\log M}{M} \right)^{1-\frac{p}{2}} \; \left(\frac{\sigma^2_{\text{max}}}{N} \right)^{1-\frac{p}{2}} \; 2^{-s'pj_0}.\notag
\end{align}
For $T_{4.2}$, we have with Cauchy-Schwartz and exponential inequalities:
\begin{align}
  T_{4.2} \; & \leq \sum_{j=j_0+1}^{j_1} \sum_k 9 \E \left( |\varepsilon^d_{\bullet jk}|^4  \right)^{\frac12} \E \left[ \left( \mathbf{1}_{|\varepsilon^d_{\bullet jk}|> \lambda \sigma_{jk}/\sqrt{MN}} \right)^2 \right]^{\frac12} \notag \\
 & C_{4.2} \leq \sum_{j=j_0+1}^{j_1} \frac{\sigma^2_{\text{max}}}{MN} \exp \left[ \frac{-\left(\lambda \sigma_{jk}/\sqrt{MN}\right)^2}{{2\sigma^2_{jk}}/{MN}}\right]^\frac12 \notag \\
 & \leq C_{4.2} \; \frac{\sigma^2_{\text{max}}}{N} M^{-2}  2^{j_1}  \leq C_{4.2} \; \frac{\sigma^2_{\text{max}}}{MN} (\log M)^{-1}.\notag
\end{align}

In order to fix the parameter $j_0$, the terms $T_3$ and $T_{4.1}$ need to be balance according to $M$, {which} leads to:
\[
 2^{j_0} = \mathcal{O} \left[ \left( \log M \right)^{\frac{1-p/2}{1+s'p}} \; (MN)^{\frac{p/2}{1+s'p}} \right]. 
\]
By replacing $2^{j_0}$ in terms $T_3$ and $T_{4.1}$, the inequality (\ref{RiskL2}) becomes:
\begin{align}
\label{RiskFinal2}
   \E \left( \| \widehat{\mu}_{N,M} - \mu \|_{2}^2  \right) & \leq C_1 \; \frac{(\log M)^{-1}}{MN}  + \frac{\sigma^2_{\text{max}}}{MN} +  C_3 \; \sigma^2_{\text{max}} \left[ \frac{\log M}{MN} \right]^{\frac{2s}{2s+1}} \left( \log M \right)^{\frac{-2s'}{2s+1}}  \notag \\  
  & \qquad \qquad + C_{4.1} \; \sigma_{\text{max}}^{2-p} \; \left[ \frac{\log M}{MN} \right]^{\frac{2s}{2s+1}} \left( \log M \right)^{\frac{-2s'}{2s+1}}\notag \\
  & \qquad \qquad \qquad \qquad + C_{4.2} \; \sigma^2_{\text{max}} \frac{M^{-\frac18}}{N \log M} + C_5 \left[ \frac{\log M}{M} \right]^{2s'} \notag 
\end{align}
The convergence of the overall expression is limited by the terms in $\mathcal{O} \left[ \left(\frac{\log M}{M} \right)^{2s'} \right]$ and in
$$\mathcal{O} \left[ \left( \frac{\log M}{MN} \right)^{\frac{2s}{2s+1}} \left( \log M \right)^{\frac{-2s'}{2s+1}} \right].$$
The latter leads to a limitation in 
\begin{align*}
 \mathcal{O} \left[ \left( \frac{\log M}{MN} \right)^{\frac{2s}{2s+1}} \right] & \qquad \text{if} \; \; \frac{2}{2s+1} < p < 2\\
 \mathcal{O} \left[ \left( \frac{1}{MN} \right)^{\frac{2s}{2s+1}} \right] & \qquad \text{if} \; \; p \geq 2 
\end{align*}

Hence, we get:
\[
 \E \left( \|\widehat{\mu}_{N,M} - \mu \|_{2}^2  \right) \leq \max \left\{ \mathcal{O}\left[ \left(\frac{\log M}{MN}\right)^{\frac{2s}{2s+1}} \right] + \left[\mathcal{O} \left( \frac{\log M}{M} \right)^{2s'} \right] \right\},
\]
that concludes the proof.

\subsection{Derivation of the SURE criterion for SCAD thresholding}
\label{appendix:SURE.SCAD}

 For recall, the SCAD thresholding function is given by:
\begin{equation}
  \label{SCADthresh}
    \delta^{\text{scad}} (d_{jk},\lambda,a) = \left\{
  \begin{array}{ll}
  \sign(d_{jk})(|d_{jk}| - \lambda)_+ & \text{ si } \; |d_{jk}| \leq 2\lambda , \\
  \frac{(a-1)d_{jk}- a \lambda \sign(d_{jk})}{a-2} &  \text{ si } \; 2\lambda < |d_{jk}| \leq a\lambda , \\
  d_{jk} &  \text{ si } \; |d_{jk}| > a\lambda.
 \end{array}
 \right.
\end{equation}
and we are looking for a function $\gbf:\rset^{2^j} \rightarrow \rset^{2^j}$ such that:
\begin{equation}
 \label{DecompSCADfn}
  \delta^{\text{scad}} (\dbf_j,\lambda,a) = \dbf_j + \gbf(\dbf_j),
\end{equation}
to define the SURE-SCAD criterion:
\begin{equation}
\label{SURE.SCAD.Cr}
 \text{SURE}_{\text{scad}} (\lambda;\dbf_j) = 2^j + \| \gbf(\dbf_j) \|_2^2 + 2 \sum_{k=0}^{2^j-1} \frac{\partial \gbf (\dbf_{jk})}{\partial {d}_{jk}},
\end{equation}
\noindent with $\gbf (\dbf_j) = \big( g_{j_0} (d_{j0}),\ldots,g_{2^j-1} (d_{j,2^j-1}) \big)$.
By defining $g$ as the weakly differentiable function:
\begin{multline*}
 g(d_{jk}) = -d_{jk} \mathbf{1}_{ \{ |d_{jk}| \leq \lambda \} } - \lambda \sign (d_{jk}) \mathbf{1}_{ \{ \lambda < |d_{jk}| \leq 2\lambda \} } \\ + \left( \frac{d_{jk}}{a-2} + \frac{a\lambda \sign (d_{jk}) }{a-2} \right) \mathbf{1}_{ \{ 2\lambda < |d_{jk}| \leq a\lambda \} }, 
\end{multline*}
with $\gbf(\dbf_j) = \big( g(d_{j0}),\ldots,g(d_{j,2^j-1}) \big)$, the relation (\ref{DecompSCADfn}) is satisfied. We can then compute:
\begin{align*}
 \| \gbf(\dbf_j) \|^2_2 & = \sum_{k=0}^{2^j-1}  g(d_{jk})^2 \\
 & \quad \text{with} \; g(d_{jk})^2 = d_{jk}^2 \mathbf{1}_{ \{ |d_{jk}| \leq \lambda \} } + \lambda^2 \sign (d_{jk}) \mathbf{1}_{ \{ \lambda < |d_{jk}| \leq 2\lambda \} } \\
 & \qquad \qquad \qquad \qquad + \frac{1}{(a-2)^2} \left[ d_{jk}^2 + (a\lambda)^2 +2a\lambda |d_{jk}| \right] \mathbf{1}_{ \{ 2\lambda < |d_{jk}| \leq a\lambda \} } \\
&  \sum_{k=0}^{2^j-1} \frac{\partial g(d_{jk})}{\partial d_{jk}} = \sum_{k=0}^{2^j-1} \left[ -\mathbf{1}_{ \{ |d_{jk}| \leq \lambda \} } + \frac{1}{a-2} \mathbf{1}_{ \{ 2\lambda < |d_{jk}| \leq a\lambda \} } \right],
\end{align*}
which leads finally to the criterion (\ref{SURE.SCAD}) in Section \ref{ThStrategies}.

\newpage

\newpage

\newpage

\newpage

\newpage
\begin{table}[ht!]
\begin{small}
\begin{center}
\begin{tabular}{|l|c|c|c|c|c|c|c|c|}
\hline
& \multicolumn{4}{|c|}{SNR $ =1$} & \multicolumn{4}{|c|}{SNR$ =5$}\\
\cline{2-9}
& \multicolumn{2}{|c|}{$\tau =0.1$} & \multicolumn{2}{|c|}{$\tau =1$} & \multicolumn{2}{|c|}{$\tau =0.1$} & \multicolumn{2}{|c|}{$\tau =1$}\\ 
\cline{2-9}
& Ho & He & Ho & He & Ho & He & Ho & He \\
 \hline
Blocks & 5.093 & 0.168 & 0.424 & 0.166 & 0.186 & 0.001 & 0.011 & 0.001 \\ 
& (1.629) & (0.018) & (0.130) & (0.017) & (0.054) & (2e-4) & (0.006) & (2e-4) \\
 \hline
Bumps & 5.028 & 0.724 & 0.944 & 0.720 & 0.220 & 0.040 & 0.0573 & 0.040 \\ 
& (0.745) & (0.025) & (0.048) & (0.027) & (0.029) & (0.001) & (0.002) & (0.001) \\ 
 \hline
Heavisine & 5.293 & 1.193 & 1.773 & 1.192 & 0.530 & 0.079 & 0.129 & 0.079 \\ 
($\times 10^{-2}$)& (0.303) & (0.103) & (0.120) & (0.104) & (0.016) & (0.006) & (0.008) & (0.006) \\  
 \hline
Doppler & 26.79 & 5.607 & 7.819 & 5.629 & 1.387 & 0.187 & 0.304 & 0.188 \\ 
($\times 10^{-4}$)& (4.058) & (2.607) & (0.320) & (0.238) & (0.136) & (0.117) & (0.015) & (0.010) \\
\hline
\end{tabular}
\end{center}
\end{small}
\caption{Average MISE (and associated standard deviations) on 200 repetitions for the fixed effects \texttt{Blocks}, \texttt{Bumps}, \texttt{Heavisine} and \texttt{Dopller} in a heteroscedastic framework. The heteroscedastic structure is defined as in equation \ref{HeteroStruct} with SNR and $\tau$ varying respectively in (1,5) and (0.1,1). The sample size is set to $N=100$ and the signal size to $M=1024$. Final estimates are based on a SCAD thresholding using the universal threshold $\lambda_U$.}
\label{Table:ExResHomoVsHetero}
\end{table}

\newpage
\begin{table}[ht!]
\begin{small}
\begin{center}
\begin{tabular}{|l|c|c|c|c|}
\hline
& \multicolumn{2}{|c|}{SNR $ =1$} & \multicolumn{2}{|c|}{SNR$ =5$}\\
\cline{2-5}
& Homoscedastic & Heteroscedastic & Homoscedastic & Heteroscedastic \\
 \hline
Blocks & 0.189 & 0.168 & 1.44e-3 & 1.43e-3  \\ 
& (0.016) & (0.017) & (2.5e-4) & (2.5e-4) \\
 \hline
Bumps & 0.736 & 0.726 & 0.045 & 0.040 \\ 
& (0.024) & (0.024) & (1.25e-3) & (1.25e-3) \\ 
 \hline
Heavisine & 1.203 & 1.204 & 0.079 & 0.078 \\ 
($\times 10^{-2}$)& (0.097) & (0.104) & (0.006) & (0.006)  \\  
 \hline
Doppler & 5.658 & 5.622 & 0.201 & 0.188  \\ 
($\times 10^{-4}$)& (0.246) & (0.274) & (0.011) & (0.011) \\
\hline
\end{tabular}
\end{center}
\end{small}
\caption{Average MISE (and associated standard deviations) on 200 repetitions for the fixed effects \texttt{Blocks}, \texttt{Bumps}, \texttt{Heavisine} and \texttt{Doppler} in a homoscedastic framework (with $\sigma^2_{jk} = \sigma^2$ for all $(j,k) \in \Lambda$). The noise level is controlled by the SNR ratio varying in (1,5). The sample size is set to $N=100$ and the signal size to $M=1024$. Final estimates are based on a SCAD thresholding using the universal threshold $\lambda_U$.}
\label{Table:ExResHomo}
\end{table}

\end{document}